\documentclass[prl,twocolumn,showpacs]{revtex4}
\usepackage{hyperref}
\def\ltsim{\vbox {\hbox{\lower .8\baselineskip \ hbox{$<$}} \break
                 \hbox{\lower 0.2\baselineskip \hbox{$\sim$}} } }

\begin{document}

\title{Double-dot charge qubit and transport via dissipative cotunneling}
\author{Mei-Rong Li and Karyn Le Hur}
\affiliation{D\'epartement de Physique, Universit\'e de Sherbrooke, 
Sherbrooke, Qu\'ebec, Canada J1K 2R1}
\date{\today} 

\begin{abstract}
We investigate transport through an exotic ``charge'' qubit composed of two 
strongly capacitively coupled quantum dots (QDs), each being independently 
connected to a side gate which in general exhibits a fluctuating electrostatic 
field ({\em i.e.}, Johnson-Nyquist noise). Two quantum phases are found: the 
``Kondo'' phase where an orbital-Kondo entanglement emerges and a ``local 
moment'' phase in which the noise destroys the Kondo effect leaving the orbital 
spin unscreened and resulting in a clear suppression of the conductance. In the 
Kondo realm, the transfer of charge across the setting is accompanied by 
zero-point charge fluctuations in the two dissipative environments and then the 
I-V characteristics are governed by what we call ``dissipative cotunneling''. 
 \end{abstract}
\pacs{73.23.-b, 72.15.Qm, 42.50.Lc}
\maketitle

The analogy between the Coulomb blockade effect in quantum dots (QDs) and the Kondo 
effect in heavy fermion systems has been emphasized for a decade \cite{Glazman}. 
Recently, fascinating quantum phase transitions were discovered in certain heavy 
fermion materials \cite{Lohneysen}, and a spin-boson-fermion impurity model has been 
proposed to capture the quantum passage from local moment magnetism to 
strongly-correlated Kondo behavior \cite{Si}. We seek to explore the emergence 
of similar physics at the mesoscopic scale through a setup composed of a double-QD 
``charge'' qubit immersed in a noisy electromagnetic environment. The central goal 
of this Letter is to focus on transport properties close to the quantum phase 
transition. We stress that a strong capacitive coupling between dots is essential 
for the fabrication of the double-QD charge qubit. Two capacitively coupled QDs have 
recently attracted much attention \cite{Andrei}. One of the main motivations certainly 
lies in the fact that having coupled qubits is required for the controlled-NOT 
operation, one of the universal gates in quantum computing \cite{Pashkin}. They are 
already known to exhibit non-trivial degeneracy points where quantum fluctuations 
can lead to exotic emergent states with orbital Kondo physics \cite{Zarand}. 
Experimentally, coupled QDs can be built up, {\em e.g.}, in semiconducting 
heterostructures \cite{WaughWeis} and in nanotubes \cite{Mason}. 

{\em Model of two strongly capacitively coupled QDs subject to quantum noises.}---The 
schematic circuit diagram of the setup under consideration is shown in 
Fig.~1. Each QD couples to a side-gate and each source of gate voltage
is placed in series with an impedance $Z_{1,2}(\omega)$ showing resistive
behavior at low frequency $\omega$. First,
we will neglect the reservoirs of electrons (left and right leads)
and seek to analyze
the Coulomb blockade effect. At the Hartree-Fock level,
the charging energy takes the form \cite{Pashkin}
\begin{figure}[h]
\begin{picture}(250,120)
\leavevmode\centering\includegraphics{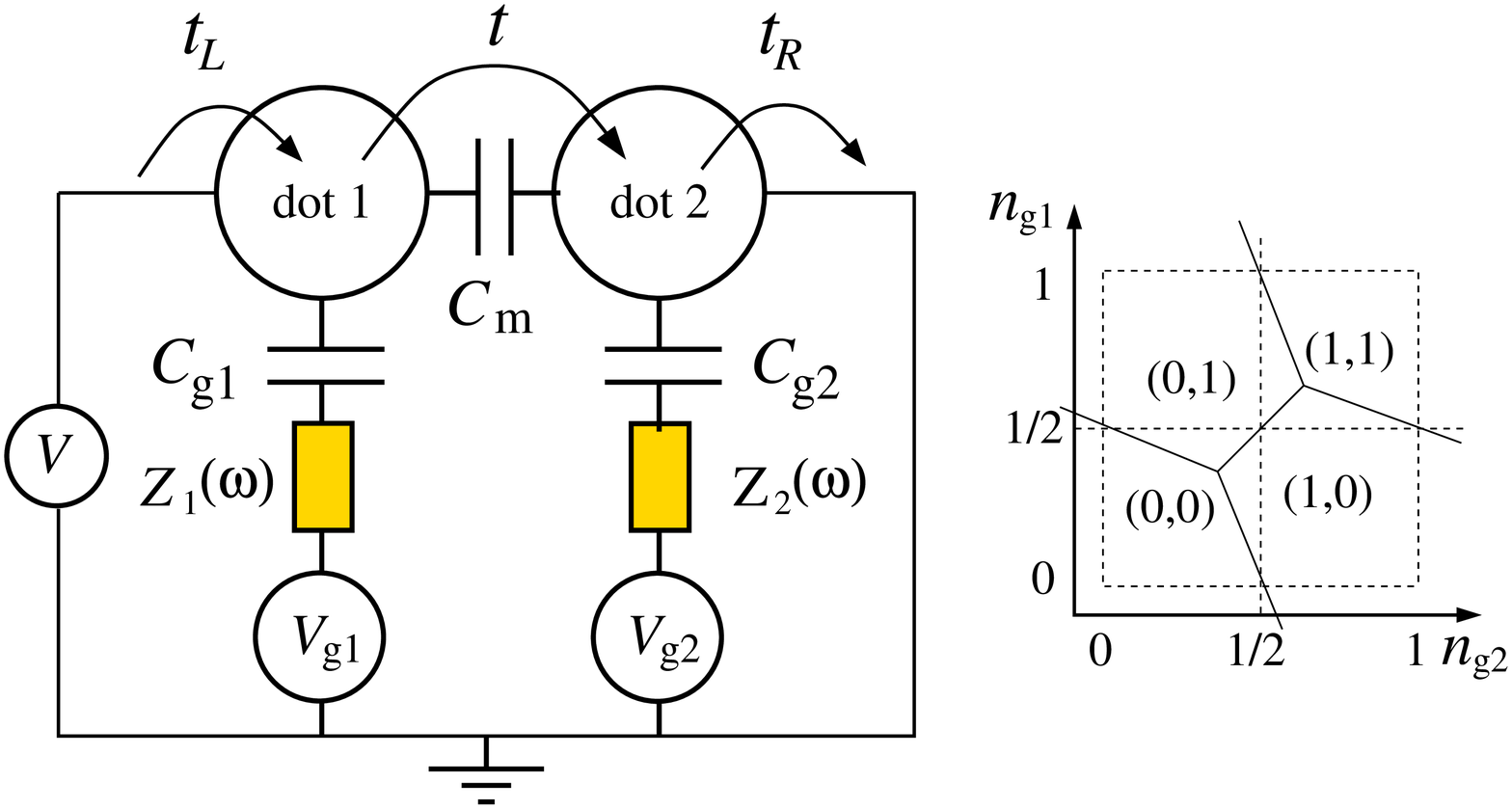}
\end{picture}
\caption{Left: Schematic setup of a double-QD charge qubit being connected in
series to reservoir leads. The two QDs are capacitively coupled to side-gates and 
fluctuations of the gate voltages are included as a result of the finite resistances 
$R_{1,2}=Z_{1,2}(\omega=0)$ in the gate leads. Right: Honeycomb pattern of the 
charging energy for $C_m\gg C_{gi}$ where $i=1,2$. Our concern is close to the 
degeneracy point $n_{g1}=n_{g2}=1/2$.} 
\end{figure}
\begin{eqnarray}
E_{n_1n_2}&=&E_{c1} \left(n_{g1}-n_1\right)^2 + E_{c2} \left(n_{g2}-n_2\right)^2 
\nonumber \\
&& + E_m \left(n_{g1}-n_1\right)\left(n_{g2}-n_2\right);  \label{chargingenergy}
\end{eqnarray}
we have $E_{c1,2}=e^2 C_{\Sigma 2,1}/2(C_{\Sigma 1}C_{\Sigma 2}-C^2_m)$, 
$E_{m}=e^2C_m/(C_{\Sigma 1}C_{\Sigma 2}-C^2_m)$, $n_{g1,2}=C_{g1,2}V_{g1,2}/e$, and 
$C_{\Sigma 1,2}=C_{g1,2}+C_m$. In Eq.~(\ref{chargingenergy}), $n_1$ and $n_2$ are 
the numbers of electrons on each dot with respect to the equilibrium state at 
$V_{g1,2}=0$. Our current interest lies in the degeneracy point $n_{g1}=n_{g2}=1/2$ 
and in the strong capacitive coupling limit $C_m\gg C_{g1,2}$, so that
$E_{01}=E_{10}\ll E_{11}=E_{00}$ with $E_{11}-E_{01}=E_m/2\simeq E_{c1}\simeq E_{c2}$. 
The low-energy physics can thus be studied within the restricted Hilbert space in 
which only the (0,1) and (1,0) states are allowed (see Fig.~1). In this case, the 
double QD effectively reduces to a charge qubit (two-level system) and one can 
introduce an orbital spin $\hat{T}_z$ to describe the qubit, with the orbital spin 
up and down representing the (0,1) and (1,0) states respectively. It is convenient 
to introduce the operators $\widehat{P}_{01}$ and $\widehat{P}_{10}$ projecting on 
the states $(0,1)$ and $(1,0)$ respectively. Through the fluctuation-dissipation 
theorem, the impedance $Z_{1,2}(\omega)$ introduces gate voltage fluctuations 
$\delta V_{g1,2}$. Since $\hat{T}_z=(\widehat{P}_{01}-\widehat{P}_{10})$, 
the Hamiltonian describing the charge fluctuations on the charge qubit due to 
$\delta V_{g1,2}$ then can be written as ($\langle \delta V_{gi}(t) \rangle=0$)
\begin{eqnarray}
H_{fl}=e\left(\delta V_{g1}-\delta V_{g2}\right) \hat{T}_z.  
\label{couplenoise}
\end{eqnarray}
We plan to consider the case of Johnson-Nyquist noise where the resulting gate 
voltage fluctuations are embodied by $\langle \delta V_{g1,2} \delta V_{g1,2}
\rangle_\omega = \hbar R_{1,2}\, \omega \coth (\hbar\omega/2k_BT)$; 
this can be generally attributed to the presence of ohmic resistances $R_{1,2}$ 
in the gate leads. Below, we model the resistances $R_{1,2}$ in a microscopic fashion 
through two long dissipative transmission lines being composed of infinite collections 
of $L_iC_{ti}$ oscillators [Here $i=1,2$ is the dot (and transmission line) index].
For a given gate lead, the resistive external circuit containing the transmission line 
has been previously introduced in Ref.~\onlinecite{lehur}. For the sake of clarity, 
here we follow Ref.~\onlinecite{Buttiker} by denoting the charge (fluctuation)
operator on the capacitor $C_{gi}$ by $\hat{Q}_{0,i}$ and the ones on the capacitors 
$C_{ti}$ between the inductances $L_i$ by $\hat{Q}_{l,i}$ with $l=1,2,...$, and 
introducing the corresponding conjugate phase operators  $\hat{\phi}_{l,i}$. 
These transmission lines produce ohmic resistance $R_i=\sqrt{L_i/C_{ti}}$ and 
are described by the bosonic Hamiltonian:
\begin{eqnarray}
H_{B}= \sum_{i=1,2}  \bigg\{ {\hat{Q}^2_{0,i}\over 2C_{gi}} + \sum^{+\infty}_{l=1} 
\bigg[{\hat{Q}^2_{l,i}\over 2C_{ti}}+
{\hbar^2\over e^2} {(\hat{\phi}_{l,i}-\hat{\phi}_{l-1,i})^2\over 2L_i} \bigg]\bigg\} 
\label{LC}.
\end{eqnarray}
To diagonalize $H_B$ analytically, we assume $C_{ti}= C_{gi}$. Using the transformation 
\cite{Buttiker} 
$\hat{Q}_{l,i}  =\sqrt{2}\int^1_0 dx \cos[(l+1/2)\pi x] \hat{Q}_i(x)$ and 
$\hat{\phi}_{l,i}=\sqrt{2}\int^1_0 dx \cos[(l+1/2)\pi x] \hat{\phi}_i(x)$, with 
$[\hat{\phi}_j(x),\hat{Q}_k(y)]=ie\delta(x-y)\delta_{jk}$, and redefining 
$\hat{Q}_i(x)$ and $\hat{\phi}_i(x)$ in terms of boson operators $\hat{a}_{xi}$, 
we have
\begin{equation}
H_B=\sum_{i=1,2}\int^1_0 dx  \, \hbar\omega_{xi} \left(\hat{a}^+_{xi} \hat{a}_{xi}
+{1\over 2}\right),
\end{equation}
where $\omega_{xi}=2\omega_{ci}\sin(\pi x/2)$ with $\omega_{ci}=\sqrt{1/L_iC_{ti}}$ 
being a realistic high-frequency cutoff. The gate voltage fluctuations 
$\delta V_{gi}=\hat{Q}_{0,i}/C_{gi}$ in Eq.~(2) read \cite{Buttiker}
\begin{eqnarray}
\delta V_{gi}=
\frac{1}{C_{gi}}\int^1_0 dx \cos\left({\pi x\over 2}\right) 
\sqrt{\hbar\omega_{xi}C_{ti}}
\left(\hat{a}_{xi}+\hat{a}^+_{xi}\right). \label{voltagefluctuation}
\end{eqnarray}
It thus becomes clear that Eq. (2) introduces a coupling between the orbital spin and 
two boson baths.

Now we incorporate electron tunneling between the two QDs. Throughout this paper we 
imply that a strong in-plane magnetic field has been applied, which allows us to 
restrict ourselves to spinless electrons. We still have to distinguish between large 
(metallic) QDs and small QDs. Large QDs are at the micron scale and then yield a very 
dense energy spectrum. In contrast, a small QD is at the nano scale and the energy 
level spacing is quite large. 

We first consider the case of large QDs. They are in the many-body state 
$|\Phi_1\Phi_2\rangle$ described by the free kinetic Hamiltonian 
$H^{(0)}_{\rm dot}=\sum_{i=1,2}\sum_{p_i} \epsilon_{p_i} d^+_{ip_i} d_{ip_i}$, 
where $d_{ip_i}$ is the annihilation operator for an electron in the state $p_i$ on 
the $i$th dot. However, the Coulomb blockade effect analyzed above dictates that the 
tunneling can only take place between the (0,1) and (1,0) states. The key point is 
that one must use the projection operators $\widehat{P}_{01}$ and $\widehat{P}_{10}$
to actually project on $|\Phi_1\Phi_2\rangle$ with charge number $n_1=0, n_2=1$ and 
$n_1=1, n_2=0$, respectively. The tunneling Hamiltonian then becomes
\begin{eqnarray}
H^{\rm (large)}_t=t\sum_{p_1p_2} d^+_{1p_1}d_{2p_2} \widehat{P}_{01}
+t^*\sum_{p_1p_2} d^+_{2p_2}d_{1p_1} \widehat{P}_{10}. \label{tunneldots1}
\end{eqnarray}
By explicitly introducing an auxiliary charge label $|n_1n_2\rangle$ 
to $|\Phi_1\Phi_2\rangle$ the Hamiltonian $H_t$ in Eq. (\ref{tunneldots1}) 
then turns into  
\begin{eqnarray}
H^{\rm (large)}_{t}=t \sum_{p_1p_2} d^+_{1p_1}d_{2p_2} \hat{T}_- +  
t^*\sum_{p_1p_2} d^+_{2p_2}d_{1p_1}\hat{T}_+.
\label{tunneldots2}
\end{eqnarray}
The tunneling matrix element $t$ is assumed not to depend on the momenta $p_1$
and $p_2$. The $\hat{T}_+$ operator -- acting exclusively on $|n_1n_2\rangle$
-- ensures that each time an electron travels from dot 1 to dot 2 the charge 
state is explicitly adjusted from $|10\rangle$ to $|01\rangle=\hat{T}_+|10\rangle$. 
This causes a flip in the orbital spin which is accompanied by particle-hole 
excitations shared between the two dots. We find it appropriate to visualize 
$H^{\rm (large)}_t$ as a transverse Kondo coupling along the lines of a 
single-electron box coupled to a reservoir lead \cite{Mateev}. The Kondo energy 
scale below which the transmission between dots becomes almost perfect reads 
$T^{\rm (large)}_{K0}=D_0 e^{-1/(2|t|N_0)}$; the ultraviolet energy cutoff is 
$D_0\sim E_{c1,2}$ and $N_0$ is the density of states on the two dots which has 
been assumed to be symmetric.

The total Hamiltonian for the two large dots in the absence of the leads can be 
summarized as 
\begin{eqnarray}
H^{\rm (large)}_{\rm dot}=H_{fl}+H_B+H^{\rm (large)}_t+H^{(0)}_{\rm dot}. 
\label{HLtotal}
 \end{eqnarray}
This is now clearly mapped onto a {\it spin-boson-fermion} model \cite{Zhu}. We 
find the following renormalization group (RG) equations: $d\lambda_\perp/dl=
\lambda_\perp\lambda_z-g\lambda_\perp/2,\ d\lambda_z/dl=\lambda_\perp^2$, and 
$dg/dl=-g\lambda_\perp^2$ with $l=\ln(D_0/D)$ denoting the scaling variable 
($D$ is the energy variable). Here, $\lambda_\perp=2|t|N_0$, $\lambda_z$ is the 
induced Ising-like coupling between the orbital spin and the fermion degrees of 
freedom, and $g$ is an important physical parameter defined as
$g=4\left(R_1+R_2\right)/R_Q$ (in the case of $C_{ti}=C_{gi})$; 
$R_Q=h/e^2\approx 25.8k\Omega$ is the quantum of resistance. Like in Ref. 
\onlinecite{lehur}, we rewrite the RG equations in terms of the variables
$\lambda_\perp=2|t|N_0$ and $\tilde{\lambda}_z=\lambda_z-g/2$; initially,
$\tilde{\lambda}_z=-g/2\leq 0$. Then, we recognize the well-known Kosterlitz-Thouless 
(KT) equation flow $d\lambda_\perp/dl= \lambda_\perp\tilde{\lambda}_z$ and 
$d\tilde{\lambda}_z/dl=\lambda_\perp^2$, which reveals a quantum phase transition 
at $\tilde{\lambda}_z=-{\lambda_\perp}$, {\em i.e.}, $g=g_{cl}=4|t|N_0$ which 
experimentally corresponds to $R_1+R_2=|t|N_0  R_Q<R_Q$ in the weak tunneling 
limit $|t|N_0\ll 1$. For $g<g_{cl}$, the charge qubit is still in the Kondo realm 
with the Kondo temperature being renormalized by noise: $T^{\rm (large)}_{K}(g)
\approx D_0 e^{-1/(2|t|N_0-g/2)}$; the boson baths have additive effects. For 
$g>g_{cl}$, the noise effect becomes dominant and eliminates the orbital Kondo effect; 
the hopping between the two QDs becomes tiny and the current is strongly suppressed
(see below).

In contrast to the large QDs, the small QDs are in a truly two-level state at 
the degeneracy point $n_{g1}=n_{g2}=1/2$ due to the large energy level spacing. 
Tunneling of electrons between the two QDs still induces an orbital spin flip 
however there is no particle-hole excitation
involved. In consequence, the tunnel Hamiltonian between the two small dots 
takes the more conventional form \cite{Zarand}
\begin{eqnarray}
H^{\rm (small)}_{\rm dot}=H_{fl}+H_B+(t\hat{T}_-+h.c.). \label{HStotal}
\end{eqnarray}
Such a model has been already investigated in great detail, including in the
context of a small noisy dot embedded in a mesoscopic ring \cite{Buttiker}. 
According to Ref.~\onlinecite{Buttiker}, this can be mapped exactly onto a 
transverse {\it spin-boson} model \cite{Leggett} which in passing has been 
demonstrated to apply to a wide range of two-level state problems including a 
superconducting qubit \cite{Schon} and a quantum dot connected to Bose-Einstein 
condensate reservoirs \cite{BEC}. The phase diagram can also be derived from 
an analogy with the single-impurity (anisotropic) Kondo model \cite{Leggett}. 
This already ensures some universality in the transport features independently 
of the precise sizes of the QDs. Here the Kondo energy scale 
$T_K^{(small)}(g)=|t|\{|t|/D_0\}^{g/(1-g)}$ only decays as a power-law with 
$g$ and that the KT phase transition should arise in the vicinity of 
$g=g_{cs}\approx 1$ \cite{Buttiker}.
  
Now, we are in a favorable position to accurately study transport properties 
through the double-QD charge qubit in the quantum limit of $T=0$. We treat the 
case of the large dots and the small dots in order.

{\em Transport through large QDs.}---We connect the two QDs in series to two 
leads (source and drain) and add a voltage drop $V$ across the two leads as shown 
in Fig.~1. In the restricted Hilbert 
space, at small energy $eV$, single-electron sequential tunneling through 
the dots is exponentially suppressed \cite{Sch}. The leading order contribution
to the tunneling here comes from the {\it cotunneling} process:
\begin{eqnarray}
H^{\rm (large)}_T=t' \sum_{k_1k_2} \sum_{p_1p_2} \left(c^\dagger_{2k_2} c_{1k_1}  
d^+_{1p_1}d_{2p_2} \hat{T}_- + h.c.\right), \label{leadshopping}
\end{eqnarray}
where 
$c_{ik_i}$ annihilates a spinless electron in the state 
$k_i$ on the left ($i=1$), right ($i=2$) lead;
$t'\simeq t_L t_R/ [E_{11}-E_{01}]=2t_L t_R/E_m$
with $t_L$ ($t_R$) being the tunneling matrix element between the left lead and
dot 1  (the right lead and dot 2).
Since $H_T^{\rm (large)}$ involves four fermion operators, this is 
irrelevant in the RG spirit and can always be treated 
perturbatively.

To compute the tunneling current, it is convenient to absorb the coupling
to the quantum noise into the tunneling term through the canonical 
transformation ${\cal U}={\rm exp} \{i\widehat{T}_z (\hat{\phi}_{0,1}-
\hat{\phi}_{0,2}) \}$,
where $\hat{\phi}_{0,i}={e\over \hbar}\int^t dt' \, \delta V_{gi}(t')dt'$.
The Hamiltonian (\ref{leadshopping}) then turns into: 
\begin{eqnarray}
&&\widetilde{H}^{\rm (large)}_T=t' \widehat{A} +h.c.,\label{leadshopping1}\\
&& \widehat{A} = \sum_{k_1k_2} \sum_{p_1p_2} 
c^\dagger_{2k_2} c_{1k_1} d^+_{1p_1}d_{2p_2} \hat{T}_- \,e^{i(\hat{\phi}_{0,1}-
\hat{\phi}_{0,2})}.
\label{tunnelingoperator}
\end{eqnarray}
By using linear response theory, the tunneling current then 
can be expressed as \cite{mahan}
\begin{eqnarray}
I_L=-2e \,|t'|^2 \, {\rm Im} \langle \widehat{A} \cdot
\widehat{A}^+\rangle_{\rm ret} (-eV),   \label{current}
\end{eqnarray}
where $\langle \cdots \rangle_{\rm ret}$ means the retarded correlation 
function computed with respect to the states of $H^{\rm (large)}_{\rm dot}$. It is 
important to note that the Hamiltonian in 
(\ref{leadshopping}) alone does not drive a current because it conserves the 
total number of electrons 
in the left lead and in the left dot \cite{Zarand}. It is
therefore necessary to have the tunneling term between the two 
QDs, $H^{\rm (large)}_t$, in order
to obtain a nonvanishing tunneling current. We immediately infer that 
in the boson phase $g>g_{cl}$, the current is negligible 
as a consequence of the suppressed tunneling between the two dots 
due to the fluctuating electromagnetic fields.

Now we focus on the Kondo regime $g<g_{cl}$. As discussed before, the Kondo coupling 
$\lambda_{\perp}(l)$ in $H^{\rm (large)}_t$ flows to the strong coupling limit resulting 
in a perfect transmission between the two dots. To the leading order, we then expect 
$\langle \sum_{p_1p_2}d^+_{1p_1}d_{2p_2} \hat{T}_- \cdot \sum_{p_3p_4} d^+_{2p_4}d_{1p_3} 
\hat{T}_+\rangle (\omega) \sim \delta(\omega)$ as a result of the formation of an 
Abrikosov-Suhl resonance at zero frequency $(\omega=0)$ which is the signature of a 
Kondo ground state, and thus
\begin{eqnarray}
&&I_L= 2\pi e|t'|^2 N_L N_R \int^\infty_{-\infty} d\xi \,\int^{\omega_{c}}_0
d\omega_1 \, \int^{\omega_c}_0 d\omega_2 \;f(\xi)  \nonumber \\
&&  \times [1-f(\xi+eV+\hbar\omega_1+\hbar\omega_2)] {\cal P}_1(\omega_1) \, 
{\cal P}_2(\omega_2).  \label{current1}
\end{eqnarray} 
Here $f(\xi)$ is the Fermi function (At low $eV$, we can restrict ourselves to 
equilibrium states \cite{Devoret,Nazarov}), $N_L$ and $N_R$ are the 
density of states in the left and right leads, and we introduce
the following Fourier transform 
\begin{eqnarray}
{\cal P}_i(\omega)&=&
\int^\infty_0 dt e^{i\omega t} \langle e^{i\hat{\phi}_{0,i}(t)}
e^{-i\hat{\phi}_{0,i}(0)}\rangle   \nonumber \\
&\simeq& [\exp(-\alpha_i\gamma_0)/\Gamma(\alpha_i)] \;\omega^{\alpha_i-1}
/\omega_c^{\alpha_i},
\;\;\; \label{bosonspectrum}
\end{eqnarray}
for $\omega\ll \omega_c$, where $\alpha_{i}=2R_i/R_Q$ (so $\alpha_{1}+\alpha_{2}
=g/2$), $\gamma_0\simeq0.5772$ is Euler's constant, and $\Gamma(x)$ the Gamma 
function. For convenience, we have assumed that $\omega_{c_1}=\omega_{c_2}=\omega_{c}$.
The quantity ${\cal P}_i(\omega_i)$ has a well-defined mesoscopic 
interpretation \cite{Sch,Devoret,Nazarov}: this 
represents the probability that an electron which tunnels onto the dot $i$ 
also creates
a charge excitation with an energy $\hbar\omega_i$ in the dissipative 
environment $i$. From Eq. (\ref{current1}) it becomes apparent that
the cotunneling of electrons across the double dot must be mediated 
through excitations in the two environmental bosonic modes; this will be 
referred to as ``dissipative cotunneling''. Inserting Eq. (\ref{bosonspectrum}) 
into Eq. (\ref{current1}) yields 
\begin{figure}[h]
\begin{picture}(250,164)
\leavevmode\centering\includegraphics{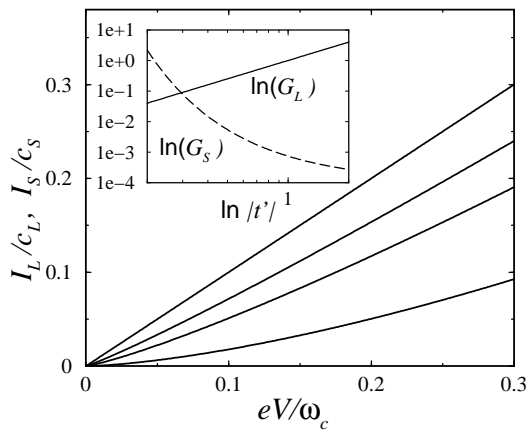}
\end{picture}
\caption{Normalized currents $I_L/c_L$ and $I_S/c_S$ as a 
function of the dimensionless bias 
voltage $eV/\omega_c$ ($\hbar=1$) in the Kondo regime. Curves from top 
to bottom correspond to $g/2=\alpha_1+\alpha_2=0,0.1,0.2,0.5$. 
The inset is a log-log plot of the conductance for large dots (solid line)
and for small dots (dashed line) versus $|t'|$ in the Kondo regime 
in the specific situation where $g=0$ and $T_{KL}\gg T_K^{(small)}(g)$ 
$(|t|/D_0=0.0001)$.} 
\end{figure}
\begin{eqnarray}
I_L=c_L [\exp(-g\gamma_0/2)/\Gamma(2+g/2)] \big(eV/\hbar\omega_{c} \big)^{1+g/2}
\label{Ilargedots}
\end{eqnarray}
at $T=0$, where $c_L=|t'|^2 N_L N_R \hbar\omega_{c}/eR_Q$. 
Again, we see that the two noises have 
additive effects. In Fig.~2, we plot  $I_L$ normalized to
$c_L$ as a function of $eV/\hbar\omega_c$ at various $g$. 
We discern that in the Kondo regime the $I-V$ 
characteristic curves of the double-QD charge qubit can be
viewed as an interesting generalization of the ones of the 
noisy tunnel junction \cite{Devoret,Nazarov} and QD \cite{Nazarov}. 
The decreasing current with growing $g$ at small voltage can be 
associated 
with the more and more suppressed noise spectrum at small $\omega$. In the 
absence of the noises, the conductance obeys $G_L=|t'|^2 N_L N_R/R_Q$ similar
to a large QD at the Coulomb blockade peaks \cite{Sch}.

{\em Transport through small QDs.}---When the two small QDs are connected in 
series to the leads, the cotunneling 
Hamiltonian becomes  $H^{\rm (small)}_T=t' \sum_{k_1k_2} 
(c^\dagger_{2k_2} c_{1k_1}\hat{T}_-+h.c.)$.
Here $H^{\rm (small)}_T$ can also generate a Kondo effect at the energy scale 
$T_{KL}=D_0e^{-1/(2|t'| N_l)}$ (we have assumed $N_L=N_R=N_l$). For quite reasonable 
$t$, in principle, we should have $T_K^{(small)}(g)\gg T_{KL}$ which guarantees
the validity of our perturbative treatment of $H^{\rm (small)}_T$, {\em i.e.}, 
Eq. (\ref{Ilargedots}) found previously for the large dots is 
then applicable. Eventually, by converging to the extreme limit of very 
small $t$, $H^{\rm (small)}_T$ could dominate the orbital Kondo physics 
and then the leading order term in the current becomes quadratic in $t$ 
\cite{Zarand}. The current at $T=0$ reads
\begin{eqnarray}
I_S=c_S [\exp(-g\gamma_0/2)/\Gamma(2+g/2)] \big(eV/\hbar\omega_{c}\big)^{1+g/2},
\label{Ismalldots}
\end{eqnarray}
where $c_S=|t/T_{KL}|^2 \hbar \omega_{c}/eR_Q$. For $g=0$, we recover the 
conductance $G_S=|t/T_{KL}|^2/R_Q$ \cite{Zarand}.
Both $I_L$ in Eq. (\ref{Ilargedots}) and $I_S$ in Eq. (\ref{Ismalldots}) have the 
same power law, {\em albeit} due to different orbital Kondo physics. A comparison 
between $G_L$ and $G_S$ versus $|t'|$ has been performed in Fig.~2 (inset). 
Note that orbital Kondo physics differs from the conventional
spin Kondo effect in which the perfect transmission usually leads to the 
conductance quantum $2/R_Q$ (the factor 2 is attributed to the two spin channels). 

In conclusion, we have carefully investigated the effect of gate voltage fluctuations 
on transport through a double-QD charge qubit. We have obtained striking 
I-V characteristics as a result of the prolific combination between orbital Kondo 
entanglement and dissipative cotunneling (the transfer of electrons through the 
structure has to be conducted by bosonic excitations in the two electromagnetic 
environments). This generalizes previous noise studies on the single tunnel junction 
and QD. When the coupling to the dissipative environments becomes large enough the 
orbital Kondo effect is suppressed and the orbital spin acts like a free moment which 
completely depletes the current (at low $V$). A similar analysis can be performed with superconducting leads and dots.

We thank L. Borda and G. Zar\'and for discussions. This work was funded by CIAR, FQRNT
 and NSERC.


\vspace{-0.6cm}


\begin{thebibliography}{10}
\bibitem{Glazman} See, {\em e.g.,} L. Kouwenhoven and L. Glazman, Phys.\ World \ {\bf 14}, 
33, (2001).
\bibitem{Lohneysen} See, {\em e.g.,} H. von L\"{o}hneysen, T. Pietrus, G. Portisch, 
H. G. Schlager, A. Schr\"oder, M. Sieck, and T. Trappmann, Phys. Rev. Lett. 
{\bf 72}, 3262 (1994). 
\bibitem{Si} See, {\em e.g.,} Q. Si, S. Rabello, K. Ingersent, and J. L. Smith,
Nature (London) {\bf 413}, 804 (2001).
\bibitem{Andrei} N. Andrei, G. Zim\'anyi and G. Sch\"on, 
Phys. Rev. B {\bf 60}, R5125 (1999).
\bibitem{Pashkin} Y. Pashkin, T. Yamamoto, O. Astafiev, Y. Nakamura, D. V. 
Averin, and J. S. Tsai, Nature {\bf 421}, 823 (2003).
\bibitem{Zarand} L. Borda, G. Zar\'and, W. Hofstetter, B. I. Halperin,
and J. von Delft, Phys. Rev. Lett. {\bf 90}, 026602 (2003).
\bibitem{WaughWeis} F. R. Waugh, M. J. Berry, D. J. Mar, R. M. Westervelt, 
K. L. Kampman, and A. C. Gossard,  Phys. Rev. Lett. {\bf 75}, 705 (1995);
U. Wilhelm and J. Weis, Physica {\bf E 6}, 668 (2000).
\bibitem{Mason} N. Mason, M. J. Biercuk, and C. M. Marcus,
  	Science {\bf 303}, 655 (2004).
\bibitem{lehur} K. Le Hur,  Phys. Rev. Lett. {\bf 92}, 196804 (2004).
\bibitem{Buttiker} P. Cedraschi and M. B\"{u}ttiker, Annals of Physics {\bf 289}, 1 (2001).
\bibitem{Mateev} K.~A. Matveev, Zh. \ Eksp. \ Teor. \ Fiz. \ {\bf 99}, 1598 (1991) 
[Sov.\ Phys. \ JETP \ {\bf 72}, 892, (1991)].
\bibitem{Zhu} L. Zhu and Q. Si, Phys. Rev. B {\bf 66}, 024426 (2002); 
G. Zar\'and and E. Demler, {\em ibid.} {\bf 66}, 024427 (2002).
\bibitem{Leggett} A. J. Leggett, S. Chakravarty, A. T. Dorsey, M. P. Fisher,
A. Garg, and W. Zwerger, Rev. Mod. Phys. {\bf 59}, 1 (1987).
\bibitem{Schon} Y. Makhlin, G. Sch\"{o}n, and A. Shnirman, Rev. Mod. Phys. {\bf 73}, 
357 (2001); 
\bibitem{BEC} A. Recati, P. O. Fedichev, W. Zwerger, J. von Delft, and P. Zoller,
cond-mat/0404533.
\bibitem{Sch} 
G. Sch\"{o}n in {\it Quantum Transport and Dissipation}, edited by T. Dittrich,
P. H\"{a}nggi, G.-L. Ingold, B. Kramer, G. Sch\"{o}n, and W. Zwerger
(WILEY-VCH, Weinheim, 1998), Chap. 3, pp. 157-159 and pp. 163-172.
\bibitem{mahan} G. D. Mahan, {\em Many-Particle Physics}, (Plenum press, New York 
and London, 1993). 
\bibitem{Devoret}
M. H. Devoret, D. Esteve, H. Grabert, G.-L. Ingold, H. Pothier and C. Urbina,
Phys. Rev. Lett. {\bf 64}, 1824 (1990).
\bibitem{Nazarov}
Yu.~V. Nazarov and  G.-L. Ingold in {\it Single Charge Tunneling}, 
edited by H. Grabert and M.~H. Devoret (Plenum
Press, New York, 1992), Chap. 2, pp. 21-107.

\end{thebibliography}
\end{document}